# Influence Process Structural Learning and the Emergence of Collective Intelligence

JAMES K. HAZY, Adelphi University and Mälardalen University, Sweden

BARAN CÜRÜKLÜ, Mälardalen University, Sweden

## 1. INTRODUCTION

Recent work [Hazy 2012] has demonstrated computationally that collectives that are organized into networks which govern the flow of resources can learn to recognize newly emerging information about the opportunities distributed in the environment. When collectives recognize opportunities as a complex adaptive system [Holland, 2001] of interacting individuals, potentially they can exploit them, but only if the individuals cooperate with respect to how resources of all kinds are allocated and utilized within the collective.

This paper argues that the system does this through a process analogous to neural network learning with relative status playing the role of synaptic weights. Hazy [2012] showed computationally that learning of this type can occur even when resource allocation decision makers have no direct visibility into the environment, have no direct understanding of the opportunity, and are not involved in their exploitation except to the extent that they evaluate the success or failure of funded projects.

Under these conditions, learning about opportunities (and threats) can occur when i) there are limited resources to allocate to copious projects each with varying outcomes, ii) projects are selected based only on the relative status of the project champions, and iii) the relative status of individuals is determined by a process that incorporates accurate feedback from the environment on the success or failure of prior projects in the environment. Effectively, the system of interactions learns which individuals have the best social networks and access to information and other resources within the ecosystem that enable them to identify opportunities (or threats) and succeed at exploiting (or defending against) them. Through trial and error, organizations can learn to exploit the diversity of their own structure.

Thus, the algorithm within the organizing network which evolves status relationships and sets decision making authority comes to the fore as an essential enabler of collective intelligence when this is defined as the capacity for a collective to recognize and exploit relevant patterns in the environment. This is particularly germane in complex organizations where no single individual or agent has access to nor fully comprehends the significance of all of the relevant information [Siggelkow and Rivkin 2005]. Hazy [2012] calls this previously unidentified emergence phenomenon: Influence Process Structural Learning (IPSL).

## 2. MODELS OF IPSL

In the prior model of IPSL, a three-tiered organizational structure was predetermined in the model design [Hazy 2012]. These initial conditions delimit the extent to which the emergence of collective intelligence can be posited. Because the model itself assumes a defined structure in the initial state, its emergence cannot be assumed. This work contributes to the field by extending the IPSL argument for collective intelligence to a holistic emergence argument.

This section begins by briefly reviewing previously published work. It continues the conversation by adding two additional steps: Firstly, it shows how a three-tier organizing structure might emerge through known complexity mechanisms. In this case the mechanism identified is preferential attachment [Barabasi 2002]. Secondly, the paper shows how collective intelligence can emerge within a system of agents when the influence structure among these agents is treated as a the genetic algorithm.

## 2.1 IPSL in a Three-Tier Organizing Structure

The model described in [Hazy 2012] assumes a decision process similar to the garbage can metaphor [Cohen, March, and Olsen 1972] with the organization arranged in three tiers. These three levels were designed in the computer model to mimic an artificial neural network [Bossomaier 2000]: input layer, hidden layer and output layer. Individual agents in the input layer find opportunities as "projects" in the environment. These are passed to certain hidden layer agents through their networks as gated by their prior relationships and relative reputations. During each time step, a limited number of projects are funded by output layer agents, "decision-makers", through a project selection procedure.

Once these projects play out over time, a back-propagation learning process adjusts the status of agents who are perceived to have advocated successful projects. In this way, the influence process structure of the organization might evolve to meet the changing needs of the ecosystem as structural variations (operationalized as status-ordering and reputation differences in relationships) are selected and retained according to improved fitness for the collective in the evolutionary sense. When this mechanism operates effectively, this system-level learning does not depend upon the learning of any individual or individuals. Rather, information about which internal network components have succeeded in the past is embedded into the local structure of the system as status and reputation asymmetries stabilize over time.

Three tiers are needed to separate input from output through the evolving structural details of a hidden layer. Within this hidden layer, IPSL operates to increase the organization's various dynamic capabilities [Helfat et al. 2007] which operate to identify and exploit opportunities in the environment [Hazy 2012]. As the organization succeeds and fails in various sub areas in a changing environment, different influence process structures emerge in specialized areas. It is posited [Hazy 2008] that these are dynamic capabilities [Helfat et al. 2007]. Information created from events as projects are implemented is gathered through feedback processes. The mechanism whereby this is back-propagated into the system as changes to some individuals' status in the organization and reputation among their peers is called the learning algorithm. These changes within the hidden layer are the means whereby a collective learns to consistently influence future interactions and move the organization further in the direction of increasing fitness.

## 2.2 The Emergence of a Three-Tier Structure

This paper contributes further specificity to the mechanisms that enable the emergence of collective intelligence. In this section, we describe a computational model that demonstrates how a three-tiered structure – with an input layer, a hidden layer and an output layer – might emerge in human interaction dynamics [Hazy and Backström 2013].

To do this, we describe an agent-based model (ABM) that is subject to moderate opportunity/threat tension [Hazy and Boyatzis, under review] that requires cooperation among agents in order to capitalize on emerging opportunities and a changing ecosystem. It is important to note that new opportunities often have previously unrecognized value, that they occur locally, and that they are first identified by individual agents. These agents are the input layer. To capitalize on these potential opportunities, however, cooperation is necessary. Agents who have identified opportunities recruit other agents in order to capitalize on opportunities, and they do this in a competitive environment. Through social network connections [Granovetter 1983] and preferential attachment mechanisms in these developing social networks [Barabasi 2002], highly centralized nodes are likely to develop as collections of agents organize into a scale-free topology [Hazy 2008].

If one assumes that the network processes information about opportunities, the most centralized nodes are posited to morph into high status individuals and eventually form an "output layer" of decision makers [Hazy 2008]. On the other hand, peripheral nodes with a limited number of edges are the "input layer". Everything in between, the intermediate nodes, can become the "hidden layer" under certain circumstances. The paper posits that this learning process emerges when the relative status and reputation differences among nodes reflect a scale-free frequency distribution as a means to efficiently process information [Baek, Bernhardsson, and Minnhagen 2011] and this relationship also reflects the relative past success of individuals within a competitive ecosystem [Hazy 2012].

### 2.3 Emergence of Collective Intelligence Through a Genetic Algorithm

This paper also contributes a computation analysis that treats influences within the three-tiered organizational structures populating the competitive ecosystem as a genetic algorithm. The model examines the evolution of influence relationships within the hidden layer (as well as the others) in the context of a fitness function as actualized on various competitive landscapes (as described by parameters). The analysis explores the conditions where emergence and adaptation, i.e. learning, occurs as a function of the nature of influence across interactions between the nodes, as well as how these nodes interact with resource conditions in the environment.

In keeping with the IPSL argument, the fitness function that is modeled is applied solely to the output layer nodes. As such, we identify conditions whereby IPSL takes place in all parts of the organizational structure (all three layers) so that decisions in the output layer improve in fitness of a collective over time. In these cases, the calculated fitness value also reflects the performance of the input and hidden layer nodes. If the benefits of fitness are equitably distributed to individuals, the emergence of collective intelligence can be recognized as a "signal" that an organization has emerged as a recognizable entity distinct from the background.

### 3. CONCLUDING THOUGHTS AND FUTURE DIRECTIONS

The arguments contained herein are supported by computational and mathematical evidence. The next step would be to frame these conclusions into falsifiable hypotheses. Many organizations would seem to exhibit what might be called a three-tier organization: executives/officers who are the decision-makers in the "output layer", middle management as the "hidden layer" and individual contributors/workers as the laboring "input layer".

Some empirical questions surfaced through the above deductive analysis: How does one find evidence of IPSL in organization? What does an adaptive learning algorithm look like? What algorithms are maladaptive? How does ISPL relate to the effectiveness of dynamic capabilities? How does ISPL relate to organizational change?

Many other questions are also possible. We suggest that all of these questions be pursued using mixed methods that combine mathematical modeling with traditional empirical techniques to continually refine hypotheses and further develop a human interaction dynamics approach [Hazy and Backström 2013] for understanding collective intelligence.